\documentstyle[sprocl]{article}

\input epsf   
\epsfverbosetrue

\def\be{\begin{equation}}
\def\ee{\end{equation}}
\def\bea{\begin{eqnarray}}
\def\eea{\end{eqnarray}}

\begin{document}

\title{THE RADIAL WAVEFUNCTIONS OF A  HEAVY-LIGHT MESON CALCULATED 
ON A LATTICE}

\author{UKQCD Collaboration, A. M. GREEN, J. KOPONEN, P. PENNANEN}

\address{Department of Physics and Helsinki Institute of Physics\\ 
P.O. Box 9, FIN--00014 University of Helsinki,Finland
\\E-mail: anthony.green@helsinki.fi, jmkopone@rock.helsinki.fi,
petrus@hip.fi} 

\author{C. MICHAEL}

\address{Department of Mathematical Sciences, University of Liverpool,
 L69 3BX, UK
\\E-mail: cmi@liv.ac.uk}

\maketitle\abstracts{A brief review is given of attempts
to understand the {\em energies} of four-quark systems calculated on
a lattice in terms of 
nuclear-physics-inspired many-body techniques involving interquark
potentials. Results are  given for the next stage of this
study where the {\em  wavefunctions}
of heavy-light mesons are also calculated on a lattice. }

Over the past few years the authors have been measuring on a lattice the
energies of various four-quark systems.
In the original papers 
(see for example Refs. \cite{GMP93} \cite{GLPM} \cite{GP98}) the four quarks
involved were all considered to be infinitely heavy. The resultant
energies could then be reasonably well understood in terms of a 
many-body nuclear-physics-inspired approach involving interquark 
potentials -- provided
there was introduced a {\em four-quark} term similar to a form factor.
Neglecting such a factor consistently led to an overestimate of the binding.

Later in Ref. \cite{MP98} a method was developed for treating on a lattice two quark
systems, where one of the quarks was a {\em light} quark i.e.  
the case of heavy-light mesons $Q\bar{q}$. In that paper the authors 
concentrated on measuring the S-, P-, D- and F-wave {\em energies}.

Returning to the four-quark system, in Refs. \cite{MP99} the energies of the 
$Q^2\bar{q}^2$ system were calculated using the same techniques that
proved successful in Ref. \cite{MP98} for the basic $Q\bar{q}$ case. In addition
to the presence of light quarks, the works of 
Refs. \cite{MP98}  \cite{MP99} had two
other improvements compared with  Refs. \cite{GMP93} \cite{GLPM}  \cite{GP98}:

\noindent i) The gauge group used was SU(3) and not SU(2).

\noindent ii) The lattice configurations were unquenched. 

In Ref. \cite{GKP} the earlier  nuclear-physics-inspired approach in terms of
interquark potentials was extended to the $Q^2\bar{q}^2$ case. This
required fitting first the $Q\bar{q}$ energies of  Ref. \cite{MP98}  to extract
an effective light-quark mass of about 400 MeV. The main conclusion from
this work was that the $Q^2\bar{q}^2$ data could not be understood in
terms of purely two-quark potentials and, as in the earlier static case
of Refs. \cite{GMP93}  \cite{GLPM} \cite{GP98}, a four-quark form factor was necessary.    
 
Most of the above work has been devoted to the $\em energies$ of the various
quark systems -- the exception being Ref. \cite{PGM} where flux-tube
structures were measured.
Now we are working on a lattice measurement of the
{\em radial wavefunctions} of
a single heavy-light meson. These wavefunctions consist of the 
distribution of the light quark and the colour field components around
the static quark. The light quark wavefunctions of the ground state and some
excited states are being
measured. Such wavefunctions have not been measured before and are of 
relevance to various phenomenological attempts to reproduce meson-decays
and scattering of mesons. These include e.g. bag models and
semirelativistic Schr\"{o}dinger and Blankenbecler-Sugar equations. 

The actual wavefunction measurement is based on the light-quark
propagators $G_{ij}$ of Ref. \cite{MP98}. 
For a measurement of the $Q\bar{q}$ energies
only one $G_{ij}$ enters in  the 2-point correlation  as, essentially,
$C_2(t)=\sum_{ij} G_{ji}U_{ij}$ where $U_{ij}$ is the static quark propagator
represented by a straight line of gauge links from point $i$ to point $j$ in
a different time slice.   
However, for the light quark wavefunction measurement two such operators arise
giving a 3-point correlation of the form
$C_3(t,r)=\sum_{ijl}  G_{jl} O G_{li}U_{ij}$,
where the site $l$ is constrained to be within $r$ spacings from the $i,j$
space coordinates.
 Here we use 
the local operators $O=\gamma_4$ and 1, which are  probing respectively the light
quark charge and matter distributions at a distance $r$ from the heavy quark.
The latter are defined as $\langle C_3(t,r)/C_2(t)\rangle$.
 The result of fitting these 
distributions by $F^2(O)$, where $F=A\exp(-r/r_0)$, is given in
Table \ref{tab:fit}. There it is seen that the charge distribution
has a considerably longer range than that of the matter.    
Summing over the charge  distribution should give the charge of 
the quark. With the present normalisation this should be $\approx$1 on
a lattice and,
within the expected accuracy, this is indeed the case, when the sum
is carried out directly on the lattice -- see the column DSum. 
As discussed in \cite{foster} the sumrule for $O=1$ has a less direct 
interpretation.
  
The authors wish to thank the Center for Scientific Computing in Finland
for their cooperation in making these studies possible.

\begin{table}[h]
\caption{ Parameters for fitting the charge ($\gamma_4$) and matter (1)
distributions with $F^2$, where $F=A\exp(-r/r_0)$. 
Dsum refers to a direct lattice estimate of the sum of $F^2$. 
\label{tab:fit}}
\vspace{0.2cm}
\begin{center}
\footnotesize
\begin{tabular}{|c|c|c|c|}
\hline
Operator($O$)  &$r_0/a$  &$A$    & DSum \\ \hline
$\gamma_4$ (Charge)&1.56(2)&0.45(1)&1.12(5)\\
1 (Matter)        &1.15(5)&0.46(2)&0.25(5)\\
\hline
\end{tabular}
\end{center}
\end{table}

\begin{figure}[b]
\includegraphics{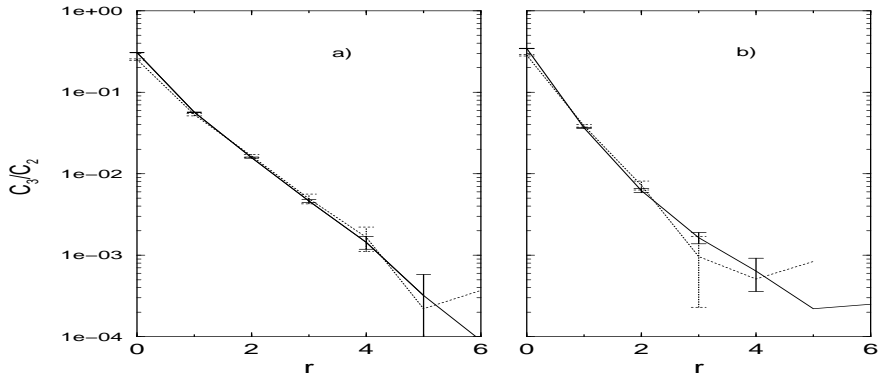}
\caption{The correlation $\langle C_3(t,r)/C_2(t) \rangle$ as a function
of $r$ in lattice units: a) Charge and b) Matter.
Solid(dotted) for $t=8(10)$.}
\end{figure}

\end{document}